\documentstyle[prl,aps,psfig]{revtex}

\begin{document}
\draft

\twocolumn[\hsize\textwidth\columnwidth\hsize\csname
@twocolumnfalse\endcsname

\widetext
\title{First-Order Pairing Transition and Single-Particle Spectral 
Function in the Attractive Hubbard Model}
\author{M. Capone, C. Castellani, and M. Grilli}  
\address{ Dipartimento di Fisica, Universit\`a di Roma ``La Sapienza'', and
INFM Center for Statistical Mechanics and Complexity, 
Piazzale Aldo Moro 2, I-00185 Roma, Italy}
\date{\today}
\maketitle

\begin{abstract}
A Dynamical Mean Field Theory analysis of the attractive Hubbard model
in the normal phase is carried out upon restricting to solutions 
 where superconducting order is not allowed.
A clear first-order pairing transition 
as a function of the coupling takes place at all the
electron densities out of half-filling between a Fermi liquid, 
stable for $U < U_c$,
and an insulating bound pairs phase for $U > U_c$, and it is 
accompanied by phase separation.
The spectral function in the metallic phase is constituted by a low energy
structure around the Fermi level, which disappears discontinuously 
at $U=U_c$,  and two high energy features 
(Hubbard bands), which persist in the insulating phase.
\end{abstract}
]

\narrowtext

The experimental finding that the (zero-temperature) coherence length 
of cuprate superconductors is much smaller than for conventional 
superconductors has suggested that these compounds lie in an intermediate
coupling regime, between the weak-coupling and the strong-coupling 
limits\cite{crossover,uemura}.
Moreover, the recent finding from angular
resolved photoemission of the existence of a (pseudo) gap in the 
single-particle spectrum well above the superconducting critical temperature,
i.e., in the normal phase,
has been usually interpreted in terms of preformed Cooper pairs
with no phase coherence.
This gave emphasis to the relevant theoretical issues related to the
description of the superconducting phase in the crossover regime
between the standard BCS and the Bose-Einstein (BE) condensation
together with the description of the normal state, where
preformed pairs or dynamical superconducting fluctuations
give rise to the pseudogap phenomenology. Regarding the
pseudogap regime, various perturbative schemes have been adopted, 
without a firm unambiguous understanding of the electron pairing
in the normal state\cite{varie}.

Much attention has been devoted to the attractive Hubbard model 
as an almost ideal framework, where the pairing between the electrons 
can be described in  all the different coupling regimes, without
complications due to other physical effects.
The Hubbard Hamiltonian reads
\begin{eqnarray}
\label{hubbard}
{\cal H} = &-&t \sum_{<ij>\sigma} c_{i\sigma}^{\dagger} c_{j\sigma} 
-U\sum_{i}\left ( n_{i\uparrow}-{1\over 2}\right )
\left ( n_{i\downarrow}-{1\over 2}\right ),
\end{eqnarray} 
where $c_{i\sigma}^{\dagger}$ ($c_{i\sigma}$) creates (destroys) 
an electron with spin $\sigma$ on the site $i$ and $n_{i\sigma} = 
c_{i\sigma}^{\dagger}c_{i\sigma}$ is the number operator; 
$t$ is the hopping amplitude and $U$ is the Hubbard on-site attraction
(we take $U > 0$, with an explicit  minus sign in the hamiltonian). 
Notice that, with this notations, the Hamiltonian is explicitly
particle-hole symmetric, so that $\mu = 0$ corresponds to 
$n=1$ (half-filling). 
Despite its simplicity, an exact solution still lacks for $d > 1$, 
and most of the known results are limited to weak ($U \ll t$)
or strong ($U \gg t$) coupling, where the BCS and the BE approaches, 
respectively, are accurate descriptions.
For $d > 2$, the ground state of the model (\ref{hubbard})
is superconducting for all values of $U$ and all densities.
At half-filling the superconducting and the  
charge-density-wave order parameters mix, due to the enlarged 
symmetry group. 

The possible formation of incoherent Cooper pairs
in the pseudogap phase of the cuprates stimulated us to disregard the
relatively well understood superconducting phase of the Hubbard model,
by constraining ourselves to solutions without superconducting order.
We rather focus on the physics of incoherent pairing
by investigating the {\it normal phase} 
within the Dynamical Mean Field Theory (DMFT)\cite{dmft,metzner}.
The DMFT is a non perturbative approach that neglects the spatial 
correlations, but fully retains the local quantum dynamics, and becomes exact
in the limit of infinite dimensions.
Due to the local nature of the interaction in the attractive Hubbard model,
we expect that the physics of local pairing is well described
(particularly in the BCS-BE crossover regime).

The existence of a {\it pairing transition} for the normal phase
at quarter filling ($n=0.5$) has been reported in Ref. \cite{keller}, where
the DMFT of the model has been performed by means of {\it finite 
temperature}  Quantum Monte Carlo (QMC) calculations.
Such a transition 
occurs between a Fermi-liquid metallic phase, and a non-Fermi liquid
phase constituted by bound electron pairs with no phase coherence.
In the same paper it was also reported a finite value of the 
quasiparticle weight $Z = (1-\partial \Sigma(\omega)/\partial \omega)^{-1}$ 
for all values of $U$, even at the pairing transition and in the pairing phase.
The relationship between this finite value of $Z$ and the spectral
properties, as well as the density dependence of the pairing 
transition, are still open questions that we address in this
paper.
We consider the Hubbard model at zero temperature
on an infinite coordination Bethe lattice of bandwidth $W$,
using the Exact Diagonalization to solve the impurity 
model\cite{caffarel}.
This method requires a truncation of the conduction bath to a small
given number of orbitals $n_s -1$, and  allows us 
to compute, directly at $T=0$, $Z$ and the density of states (DOS)
$\rho(\omega) = -1/\pi Im G(\omega)$.
A first characterization of the pairing transition may be given 
by noting that, on a bipartite lattice,
a particle-hole transformation on the down spins
$c_{i\downarrow} \to (-1)^i {c}_{i\downarrow}^\dagger$,
leaving the up spins unchanged, maps the attractive model with 
a finite density $n$ onto a half-filled repulsive model with a 
finite magnetization $m = n-1$. The 
chemical potential becomes, accordingly, a magnetic field $h = \mu$. 
In the $n=1$ case (half-filling) the two models are completely equivalent.
This mapping proves useful, since many known results for the
repulsive model and for the Mott-Hubbard transition 
can be used to gain insight on the attractive model.
In light of this mapping, the pairing transition may be
viewed as the natural counterpart of the Mott-Hubbard transition
in the presence of an external magnetic field.
Within this analogy, the normal state results that we present can be 
regarded as representative
of the physics of the attractive Hubbard model at $T > T_c$,
and eventually even provide the actual low temperature behavior, if
some mechanism frustrating superconductivity is  
effective, just like the paramagnetic solutions of the repulsive model
become relevant if frustration prevents antiferromagnetic ordering.

\begin{figure}
\centerline{\psfig{bbllx=80pt,bblly=140pt,bburx=510pt,bbury=640pt,%
figure=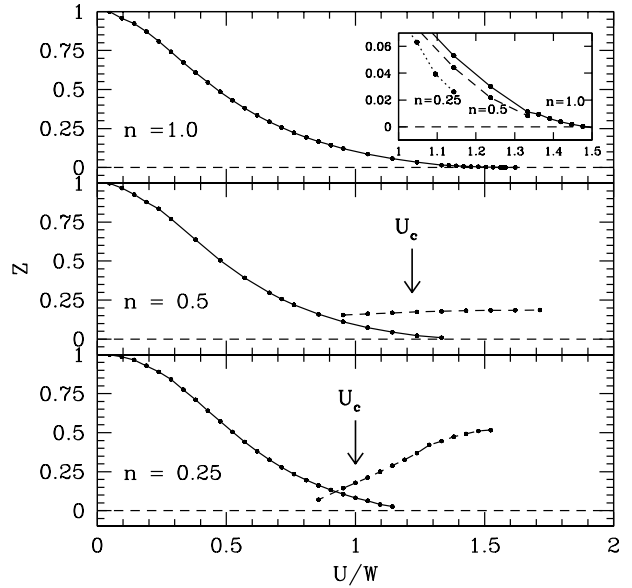,width=70mm,angle=0}}
\caption{The quasiparticle weight $Z$ as a function of $U$ for 
$n=1$, $n=0.5$, and $n=0.25$.
The solid and dashed lines join the solutions in the Fermi liquid and
pairing phase, respectively. In the half-filled 
case the latter phase has always $Z=0$.
The first-order pairing transition is marked by a vertical arrow 
(for $n=1$ the transition is second order and
coincides with the vanishing of the metallic solution $U_{c2}$).
In the inset, the metallic solutions in the proximity of their 
disappearance point $U_{c2}$ are shown. The $n=0.25$ (dotted line) and
the $n=0.5$ (dashed line) solutions have always non vanishing $Z$, while
the $n=1$ solution (solid line) vanishes at $U_{c2}$.
\label{fig_z}
}
\end{figure}

The evolution of $Z$ as a function of
$U$ for $n=1$, $n=0.5$, and $n=0.25$ is shown in Fig. \ref{fig_z}.
The results reported here are given by a linear extrapolation 
in $1/n_s$ using  
$n_s = 8, 10, 12$. In the half-filled case, the paring
state (that here coincides with the Mott insulating state) has always $Z=0$,
and $Z$ vanishes continuously  at the transition point 
$U/W=U_{c2}/W \simeq 1.49$, as reported in many previous 
studies\cite{previous,dmft}.
The numerical value agrees very well with, e.g., the recent numerical
renormalization group results of Ref. \cite{bulla}.
Away from half-filling, the metallic solution exists for
all values of $U < U_{c2}(n) < U_{c2}(n=1)$. In this phase
$Z$ is a decreasing function of $U$, but it stays 
finite for all couplings.
In particular, the disappearance of the metallic solution
at $U_{c2}$ is not associated to a vanishing 
$Z$ (see inset of Fig. \ref{fig_z}).
The pairing phase solution
exists in turn for $U > U_{c1}$, with $U_{c1}(n) < U_{c2}(n)$, and
it also has always a finite $Z$. 
In the pairing phase $Z$ is an increasing function,  
converging to the atomic limit value for large $U$.
In the interval between  $U_{c1}(n)$ and $U_{c2}(n)$, the metallic and 
pairing solutions coexist.
For the half-filled model it is known that the metallic 
solution is always energetically favored in the whole coexistence range, 
and that the two  solutions become identical  at $U_{c2}$,
where the {\it second-order} pairing transition occurs\cite{moeller}.
Away from  half-filling, the actual {\it first-order} 
transition  occurs for an intermediate coupling $U_c$ 
($U_{c1} < U_c < U_{c2}$), when the energy of the pairing state 
becomes lower than the metallic one.
The value of $U_c$  (marked by vertical arrows in Fig. \ref{fig_z}) 
is maximum in the special  half-filling case and decreases with increasing 
doping. In the extreme dilute limit $n \to 0$, the pairing transition
coincides with the binding of two electrons, and it occurs 
for $U_{c0} = 0.56W$.
We emphasize that, in general, $U_c$ has no relationship with 
the point in which the $Z$'s of the two solutions coincide,
so that $Z$ has a jump at the pairing transition.
The above results give strong evidence for the finiteness of $Z$
away from half-filling. Nevertheless, extrapolating the QMC results 
of Ref. \cite{keller} down to $T=0$, one would obtain, for $U$ close to $U_c$,
values of $Z$ significantly larger than the exact diagonalization
results reported here.
The discrepancy is easily attributed to the relatively large 
temperatures used to extrapolate the $T=0$ value. 
Further QMC calculations performed
at lower temperatures indeed show a significantly smaller
value of $Z$, which are in closer agreement with our values\cite{private}. 

The finiteness of $Z$ is a  naively surprising results, since $Z$ is 
usually interpreted as  a sort of order parameter for the Mott 
metal-insulator transition at half-filling. 
The half-filled case is, however, peculiar. In the general $n\neq 1$
case, the mapping onto the half-filled repulsive Hubbard model at finite 
magnetization, $m \neq 0$, indicates that in the Fermi liquid
$Z$ should stay finite because the low-energy
(Kondo-like) resonance characterizing the metallic state cannot
have a vanishing width due to the presence of a finite 
magnetic field $h$\cite{COSTI}. 
Also the Mott-insulating phase is different
in the presence of a magnetic field. When $h=0$
one has a pure Mott insulator with $Z=0$, whereas, when
$h$ is large enough to align {\it all} the spins,
one has a completely filled uncorrelated band for, e.g., the up spins 
and one recovers the free-electron value $Z=1$ \cite{laloux}. 
It is then natural that, at intermediate values of the magnetic field,
when $m \neq 1$ (i.e., at intermediate fillings in the attractive 
Hubbard model), $Z$ assumes finite values. 
A further insight can be given by the atomic limit ($t=0$), 
that well describes the strong-coupling limit $U \gg t$.
At half-filling,  
$\Sigma(\omega)$ diverges as $1/\omega$ 
for $\omega\to 0$, leading to $Z = 0$. 
On the other hand, away from half-filling,
the self-energy does not diverge at $\omega=0$ and $Z$ is always finite.
\begin{figure}
\centerline{\psfig{bbllx=80pt,bblly=190pt,bburx=510pt,bbury=605pt,%
figure=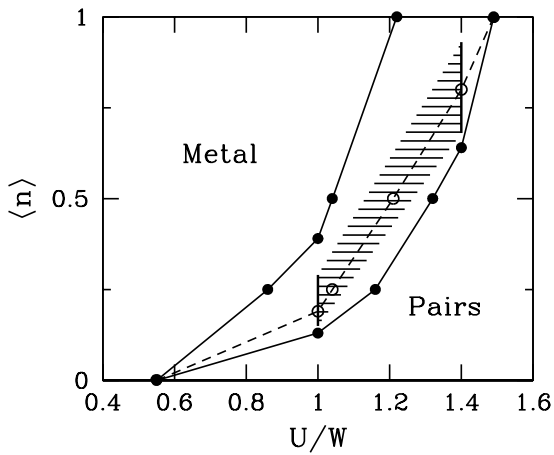,width=70mm,angle=0}}
\caption{Phase diagram in the $U$-$n$ plane. The full dots
are the calculated values of $U_{c1}$ and $U_{c2}$, while the open
dots are the pairing transition points. The thick vertical lines
mark calculated phase separation intervals for $U/W = 1$ and $1.4$,
while the shaded area is a guide to the eyes.
\label{phasediagram}
}
\end{figure}         
We now turn to the pairing transition in the grand canonical ensemble, 
where  $n$ is not fixed.
We can divide the phase diagram in the $U$-$n$ plane in four regions,
as shown in Fig. \ref{phasediagram}:
(a) $U < U_{c0} = 0.56W$,in which only the Fermi liquid solution exists
for any density; (b) $U_{c0} < U < U_{c1}$, where the metallic solution
only exists for densities from half-filling to some intermediate value, and
the insulating one exists only for small densities, and a coexistence region 
appears; (c) $U_{c1} < U < U_{c2}$, where the two solutions coexist at 
half-filling and in an adjacent region, and the metallic solution 
disappears at some density;(d) $U > U_{c2}$, where only the insulating
solution is present.
In order to reveal a possible phase separation close to the pairing
transition, we computed the density as a function of the 
chemical potential for various values of $U$. 
Phase separation occurs as soon as, for some
range of densities, the density is not an increasing function of
the chemical potential. If this is the case, a Maxwell construction
determines the phase separation region, i.e., the densities
of the phases in which the system separates.
The results are shown in the schematic phase diagram of 
Fig. \ref{phasediagram}.
In both the regions (a) and (d), as well as in the extreme point
$U = U_{c0}$, the single phase is always stable with respect to phase 
separation. 
On the other hand, in the intermediate slices (b) and (c) of the diagram, 
the first-order phase transition is always accompanied by phase separation 
between two phases at intermediate densities. 
The system therefore displays the spatial coexistence of metallic and 
insulating domains at different  densities in a finite region of 
densities around the pairing transition.

The existence of a pairing transition, its first-order character,
and the finite value of $Z$ could be expected on the basis of 
the known results for the repulsive model\cite{laloux}.
Nonetheless, the nature of the pairing phase and the mechanism leading to the
disappearance of the Fermi Liquid are less understood.
The last part of this work is therefore devoted to the 
analysis of the evolution of the DOS
as a  function of $U$ for fixed density, concentrating
on the formation of the lower and upper Hubbard bands and
on the disappearance of the quasiparticle Kondo resonance going from 
the metallic to the insulating solution.

\begin{figure}
\centerline{\psfig{bbllx=80pt,bblly=140pt,bburx=510pt,bbury=665pt,%
figure=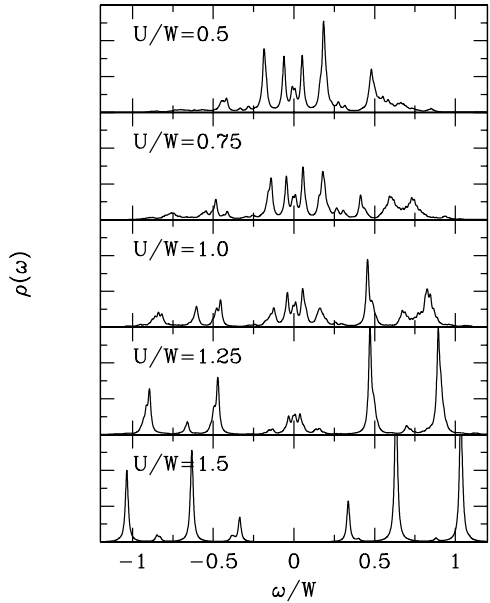,width=70mm,angle=0}}
\caption{Spectral density $\rho (\omega )$ for $n=0.75$ and various values
of $U/W$ in the metallic phase (first four panels) and in the pairing 
phase (bottom panel).
\label{fig_0.75}
}
\end{figure}

For $U=0$, the DOS is obviously the semicircular one, characteristic of
a Bethe lattice, and the chemical potential moves inside this band 
to give the desired density.
In the opposite atomic limit, we expect an insulating DOS with the
broad upper and lower Hubbard bands. Since we work out of half-filling,
the two bands will have different weight.
The effect of the attraction between the electrons is shown in Fig.
\ref{fig_0.75} for the case $n=0.75$.
Starting from small values of $U$, the first visible effect of the
interaction is a broadening of the whole spectrum,
with the high energy tails (top and bottom of the bands) moving
away from the chemical potential.
On the other hand, the total weight close to the Fermi level does not
change much increasing $U$.
Further increasing $U$, the effect is enhanced and the high-energy
weight starts to separate from the low-energy feature.
As a result, the featureless non-interacting DOS evolves into a 
well structured function, that resembles the well known result 
for $n=1$, with three distinct features: A structure around
the Fermi level, and two high-energy features, analogous
to the Hubbard bands.
Since we have $n < 1$, the upper band has larger weight than the
lower.
Regions with significant depletion of spectral weight clearly 
separate the different features.
Following the metallic solution beyond the transition point $U_c$,
a finite spectral  weight at the Fermi level is found for
all coupling values up to $U_{c2}$, where the metal abruptly disappears.
The break-down of the Fermi Liquid is not associated with
a vanishing width of the Fermi level resonance, consistently with the
results for the quasiparticle weight $Z$.
The finite $Z$ in the metallic solution is therefore associated
with this quasiparticle feature at the Fermi level.
As reported above, at $U = U_{c} < U_{c2}$, 
the metallic solution becomes energetically 
unfavored with respect to the  pairing state.
In that state the spectral function only displays the broad Hubbard bands,
and is always gapped (although  $Z$ is finite).
A similar behavior is present also for lower densities like 
$n=0.5$ and $n=0.25$ (not shown), where all
the transitions (disappearance of Fermi liquid and first-order
pairing transition) move to lower $U$, and the interval in
which three features coexist  is narrower, but clearly
present.
In all cases, even if a loss of total spectral weight close to the Fermi 
level occurs,  this weight never vanishes in the metallic phase  
and the transition to the pairing state is first order.

In conclusion, we have presented a complete characterization of the
pairing transition in the normal phase of the attractive Hubbard model.
For all densities $n \neq 1$, 
the pairing transition  is intrinsically a first-order one,
accompanied by a region of phase separation between a metallic
and an insulating phase at different densities.
The quasiparticle weight $Z$ is always finite,
and takes its minimum value at the transition point, where it
jumps from a lower metallic value to a higher value in the
pairing phase.
Even following the metallic solution in the metastability region $U_c < U \le
U_{c2}$, $Z$ never vanishes.

An analysis of the spectral function $\rho(\omega)$  shows that the finite $Z$ 
in the metallic phase is associated with a quasiparticle
peak at the Fermi level.
The evolution of $\rho(\omega)$ is quite similar
to the half-filled repulsive Hubbard model.
A structure at the Fermi level is found all the way to the
pairing transition, and two broad Hubbard bands develop and coexist with
the Fermi-level feature. 
The neglect of superconducting
symmetry breaking and the local nature of the pairing
suggest that some care must be taken in carrying over our  results 
to characterize the normal phase above $T_c$, 
particularly in systems (like cuprates) where
the pairing has a non trivial momentum structure.
However, the above DMFT analysis provides interesting indications in two
main regards. 
First of all, it shows that, for U $\simeq$ W, preformed Cooper pairs and
substantial pseudogap features can be obtained on a local basis (i.e.,
involving all momenta) even without invoking strong critical (i.e.,
at small $q$'s) pair fluctuations in the proximity of a superconducting phase
transition. This can even give rise to a phase of incoherent pairs, 
which, however,
does not seem to be generically observed in the cuprates.
At most one could argue that strongly bound incoherent
pairs are formed near the $(\pm \pi,0)$ and 
$(0,\pm \pi)$ points of the Brillouin zone, 
supporting a two-gap model for cuprates\cite{twogap}.
A second relevant outcome of DMFT analysis is the presence 
of coherent quasiparticles with a strong mass enhancement at intermediate 
coupling directly arising from the pairing interaction and coexisting with 
the high-energy pseudogap features. 
While the Hubbard-like high-energy features will survive the
turning on of coherent pairing and will be present in the normal phase 
above $T_c$, the persistence of the relatively heavy quasiparticles 
above $T_c$ is a more subtle issue.
The renormalized quasiparticles are expected to survive to superconductivity
only when $T_c$ is less than their effective bandwidth, which is  
of the order $ZW$. This could leave a very narrow window 
around $U \simeq W$ where large pairing-induced mass enhancement is visible.
 Accordingly, by crossing $U \simeq W$, the
superconducting transition will fastly evolve from a BCS-like 
instability of a Fermi Liquid (with possibly critical pair fluctuation 
induced pseudogap on an energy scale less than $ZW$) to BE-like condensation 
of preformed pairs.

We acknowledge useful discussions with W. Metzner, M. Keller, 
M. Fabrizio, C. Di Castro, and P. Pieri.
M. C. gratefully acknowledges the warm hospitality of the RWTH in Aachen.
This work is supported by MIUR Cofin 2001, prot. 2001023848 and INFM PA-G0-4.

\end{document}